\documentclass[a4paper,11pt]{article}
\usepackage{pos}

\usepackage[caption=false]{subfig}
\usepackage{cancel}

\title{QCD effects in lepton angular distributions of Drell-Yan/$Z$ production and jet discrimination}

\ShortTitle{Angular distributions of Drell-Yan/$Z$ production}

\author*[a]{Wen-Chen Chang}
\author[b,c]{Randall Evan McClellan}
\author[b]{Jen-Chieh Peng}
\author[d,e]{Oleg Teryaev}

\affiliation[a]{Institute of Physics, Academia Sinica, Taipei 11529, Taiwan}

\affiliation[b]{Department of Physics, University of Illinois at Urbana-Champaign, Urbana, Illinois 61801, USA}

\affiliation[c]{Department of Physical Sciences, Pensacola State College, Pensacola, FL 32504, USA}

\affiliation[d]{Joint Institute for Nuclear Research, 141980 Dubna, Russia}
\affiliation[e]{Dubna International University, 141982 Dubna, Russia}

\emailAdd{changwc@phys.sinica.edu.tw}
\emailAdd{remcclellan@pensacolastate.edu}
\emailAdd{jcpeng@illinois.edu}
\emailAdd{teryaev@jinr.ru}

\abstract{We present a comparison of data of lepton angular
  distributions of Drell-Yan/$Z$ production with the fixed-order pQCD
  calculations by which the baseline of pQCD effects is
  illustrated. As for the $Z$ production, we predict that $A_0$ and
  $A_2$ for $Z$ plus single gluon-jet events are very different from
  that of $Z$ plus single quark-jet events, allowing a new
  experimental tool for checking various algorithms which attempt to
  discriminate quark jets from gluon jets. Using an intuitive
  geometric approach, we show that the violation of the Lam-Tung
  relation, appearing at large transverse-momentum region, is
  attributed to the presence of a non-coplanarity effect. This
  interpretation is consistent with the appearance of violation beyond
  LO-QCD effect.}

\FullConference{%
  40th International Conference on High Energy physics - ICHEP2020\\
  July 28 - August 6, 2020\\
  Prague, Czech Republic (virtual meeting)
}

\begin{document}
\maketitle

\section{Introduction}

Measuring lepton angular distributions of Drell-Yan (D-Y)
process~\cite{peng14} provides a powerful tool to explore the reaction
mechanisms and the parton distributions of colliding hadrons. For
example, the Lam-Tung (L-T) relation~\cite{lam80} has been proposed as
a benchmark of the perturbative QCD (pQCD) effect in D-Y
process. Violations of L-T relation were observed in the measurements
of D-Y production by the fixed-target experiments as well as
$\gamma^*$/$Z$ production by the CMS~\cite{cms} and ATLAS~\cite{atlas}
experiments at LHC. It is important to understand the origin of these
violations.

It is found that the violation of the L-T relation seen in CMS and
ATLAS data in the region of transverse momentum ($q_T$) greater than 5
GeV could be well described taking into account NNLO pQCD
effect~\cite{Gauld:2017tww}. On the other hand, the agreement is not
as good in a similar comparison~\cite{Lambertsen:2016wgj,chang18} for
the fixed-target data of NA10~\cite{falciano86}, E615~\cite{conway}
and E866~\cite{e866} at $q_T<3$ GeV. Transverse-momentum dependent
Boer-Mulders functions~\cite{boer99}, correlating the nucleon spin
with the intrinsic transverse momentum of partons, have been suggested
to account for a violation of the L-T relation observed at small $q_T$
in the fixed-target experiments.

In this proceedings, we show that the $q_T$ dependence of the angular
distribution coefficients, as well as the violation of the Lam-Tung
violation, could be obtained if the angular distribution coefficients
were analyzed as a function of the number of accompanying jets in
$Z$-boson production measured by the CMS and ATLAS
Collaborations~\cite{peng19}. Furthermore we compare the data of
dilepton angular parameters $\lambda$, $\mu$, $\nu$ and the L-T
violation quantity $1-\lambda-2\nu$ measured by E615~\cite{conway}
with the fixed-order pQCD calculations. Finally we interpret some
notable features of pQCD results using the geometric
model~\cite{peng16,chang17,peng18}. More results and greater details
can be found in Ref.~\cite{chang18, peng19}.

\section{Lepton angular distributions of $Z$ production and jet discrimination}

The lepton angular distribution in the $Z$ rest frame can be expressed
as~\cite{cms,atlas}
\begin{eqnarray}
\frac{d\sigma}{d\Omega} & \propto & (1+\cos^2\theta)+\frac{A_0}{2}
(1-3\cos^2\theta)+A_1 \sin 2 \theta\cos\phi + \frac{A_2}{2} \sin^2\theta \cos 2 \phi\nonumber \\
& + & A_3 \sin\theta \cos\phi + A_4 \cos\theta
+ A_5 \sin^2\theta \sin 2\phi
+ A_6 \sin 2\theta \sin\phi
+ A_7 \sin\theta \sin\phi,
\label{eq:eq1}
\end{eqnarray}
where $\theta$ and $\phi$ are the polar and azimuthal angles of
leptons in the rest frame of $Z$. Since the intrinsic transverse
momenta of the annihilating quark and antiquark is neglected in the
original Drell-Yan model, the angular distribution is simply
$1+\cos^2\theta$ and all angular distribution coefficients, $A_i$,
vanish. For a non-zero dilepton transverse momentum, $q_T$, these
coefficients can deviate from zero due to QCD effects. However, it was
predicted that the coefficients $A_0$ and $A_2$ should remain
identical, $A_0 = A_2$, i.e. the Lam-Tung relation~\cite{lam80}.

\begin{figure}[htbp]
\centering
\includegraphics[width=0.9\linewidth]{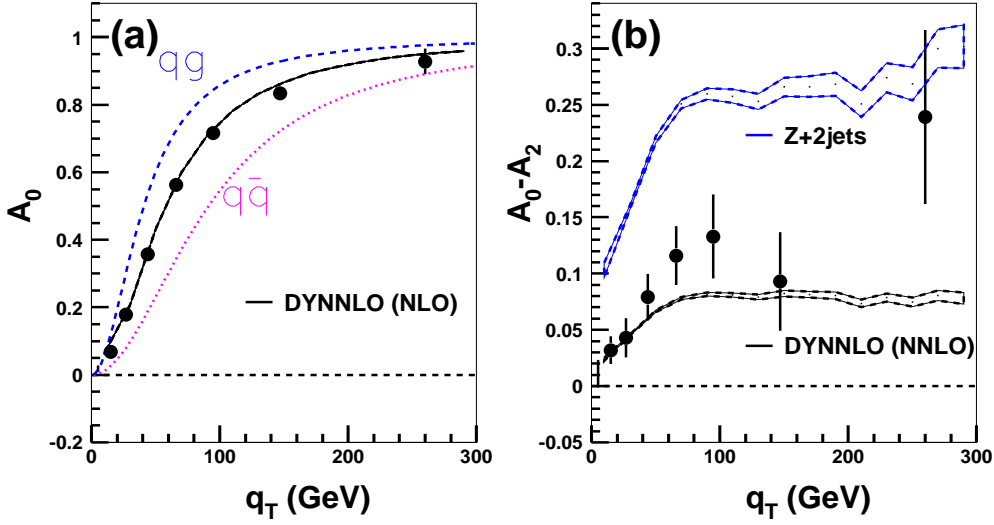}
\caption{Comparison between the CMS data~\cite{cms} of $A_0$ and $A_0
  - A_2$ for $Z$ production from $p-p$ collisions with fixed-order
  pQCD calculations. Curves correspond to calculations described in
  the text.}
\label{fig1}
\end{figure}

Figure~\ref{fig1} shows the CMS data for $A_0$ and $A_0 -
A_2$. Pronounced $q_T$ dependence of $A_0$ is observed and the
Lam-Tung relation, $A_0 - A_2 = 0$, is clearly violated. There are two
NLO QCD subprocesses for $Z$ production: $q \bar q \rightarrow Z g$
annihilation process, and $qg\rightarrow Z q$ quark Compton scattering
process. In the Collins-Soper frame~\cite{cs}, the NLO pQCD
predictions of $A_0$ and $A_2$ as a function of $q_T$ of $Z$ for these
two processes are $A_0 = A_2 = q^2_T / (M_Z^2 + q^2_T)$~\cite{collins}
and $A_0 = A_2 = 5 q^2_T / (M_Z^2 + 5q^2_T)$~\cite{thews,lindfors},
respectively. The dotted and dashed curves in Fig.~\ref{fig1}(a)
correspond to these NLO expressions.

As the $q \bar q$ and $qg$ processes contribute to the $p p \to Z X$
reaction incoherently, the observed $q_T$ dependence of $A_0$ reflects
the combined effect of these two contributions. A best-fit to the CMS
$A_0$ data, shown as the solid line in Fig.~\ref{fig1}(a), gives a
mixture of 58.5$\pm$1.6\% $qg$ and 41.5$\pm$1.6\% $q \bar q$
processes. For $pp$ collisions at the LHC, the $qg$ process is
expected to be more important than the $q \bar q$ process, in
agreement with the best-fit result. For $Z$ plus single-jet events,
Fig.~\ref{fig1}(a) shows that there is remarkable difference in the
$q_T$ dependence for $A_0$ between the $q \bar q$ annihilation process
and the $qg$ Compton process. Since it is a high-$p_T$ gluon (quark)
jet associated with the $q \bar q (qg)$ process at the $\alpha_s$
level, one could first utilize the existing algorithms for quark
(gluon) jet identification to separate the $q \bar q$ annihilation
events from the $qg$ Compton events and investigate the angular
distribution of individual event samples. Their angular distribution
coefficients for $Z$ plus single jet data would also provide a
powerful tool for testing various algorithms designed to distinguish
quark jets from gluon jets.

For the $Z$ plus multi-jet data, the L-T relation is expected to be
violated at a higher level than that of the inclusive production
data. Exclusion of the $Z$ plus single-jet events satisfying the L-T
relation, would enhance the violation of the L-T relation. We have
carried out pQCD calculations using DYNNLO~\cite{catani07,catani09} to
demonstrate this effect. Figure~\ref{fig1}(b) compares the DYNNLO
calculations with the CMS $A_0 - A_2$ data. The black band corresponds
to the NNLO calculation including contributions from the events of $Z$
with single jet and two jets. The blue band singles out the
contributions to $A_0 - A_2$ from $Z$ plus two jets only, showing that
the violation of the Lam-Tung relation is indeed amplified for the
multi-jet events. This can be readily tested with the data collected
at the LHC.

\section{Lepton angular distributions of Drell-Yan production in fixed-target experiments}

In the rest frame of the virtual photon in the D-Y process, another
expression for the lepton angular distributions commonly used by the
fixed-target experiments is given as~\cite{lam80}
\begin{equation}
\frac{d\sigma}{d\Omega} \propto 1+ \lambda \cos^2\theta
+ \mu \sin 2 \theta\cos\phi
+ \frac{\nu}{2} \sin^2\theta \cos 2 \phi,
\label{eq:eq2}
\end{equation}
where $\theta$ and $\phi$ refer to the polar and azimuthal angles of
leptons. The $\lambda, \mu, \nu$ are related to $A_0, A_1, A_2$ in
Eq.~(\ref{eq:eq1}) via
\begin{eqnarray}
\lambda = \frac{2-3A_0}{2+A_0};~~~ 
\mu  =  \frac{2A_1}{2+A_0};~~~
\nu  =  \frac{2A_2}{2+A_0}.
\label{eq:eq3}
\end{eqnarray}
Equation~(\ref{eq:eq3}) shows that the L-T relation, $1-\lambda - 2
\nu=0$, is equivalent to $A_0 = A_2$.

\begin{figure}[htbp]
\centering
\includegraphics[width=0.8\columnwidth]{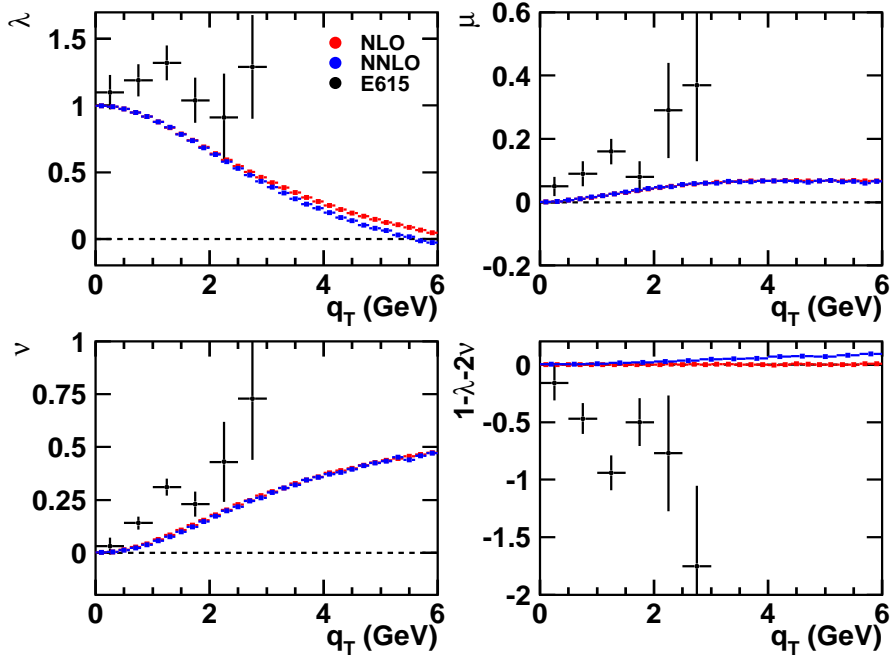}
\caption{Comparison of NLO (red points) and NNLO (blue points)
  fixed-order pQCD calculations with the E615 $\pi^-+W$ D-Y data at 252
  GeV~\cite{conway} (black points) for $\lambda$, $\mu$, $\nu$ and
  $1-\lambda-2\nu$.}
\label{fig2}
\end{figure}

In Fig.~\ref{fig2}, we compare the results of $\lambda$, $\mu$,
$\nu$, and the L-T violation, $1-\lambda-2\nu$, from the fixed-order
pQCD calculations with 252-GeV $\pi^- + W$ data from E615
experiment~\cite{conway}. The angular parameters are evaluated as a
function of $q_T$ in the Collins-Soper frame. Overall, the calculated
$\lambda$, $\mu$ and $\nu$ exhibit distinct $q_T$ dependencies.

At $q_T \rightarrow 0$, $\lambda$, $\mu$ and $\nu$ approach the values
predicted by the collinear parton model: $\lambda = 1$ and $\mu = \nu
=0$. As $q_T$ increases, Fig.~\ref{fig2} shows that $\lambda$
decreases toward its large-$q_T$ limit of $-1/3$ while $\nu$ increases
toward $2/3$, for both $q\bar{q}$ and $qG$
processes~\cite{thews,lindfors}. The $q_T$ dependence of $\mu$ is
relatively mild compared to $\lambda$ and $\nu$. This is understood as
a result of some cancellation effect, to be discussed in
Sec.~\ref{sec:discussion}.

Comparing the results of the NLO with the NNLO calculations, $\lambda
{\rm (NNLO)}$ is smaller than $\lambda \rm{(NLO)}$ while $\mu$ and
$\nu$ are very similar for NLO and NNLO. The amount of L-T violation,
$1-\lambda-2\nu$, is zero in the NLO calculation, and nonzero and
positive in the NNLO calculation. As seen in Fig.~\ref{fig2}, the pQCD
predicts a sizable magnitude for $\nu$, comparable to the data. Such
pQCD effect should be included in the determination of nonperturbative
Boer-Mulders effect from the data of $\nu$.

\section{Geometric model}
\label{sec:discussion}

As introduced above, both CMS and E615 data of lepton angular
distributions for $Z$ and D-Y production can be reasonably well
described by the NLO and NNLO pQCD calculations.  It is interesting to
see that various salient features of pQCD calculations could be
understood using a geometric approach developed in
Refs.~\cite{peng16,chang17}.

In the Collins-Soper $\gamma^*/Z$ rest frame, the hadron plane, the
quark plane, and the lepton plane of collision geometry are
defined~\cite{peng16,chang17}. A pair of collinear $q$ and $\bar q$
with equal momenta annihilate into a $\gamma^*/Z$. The momentum unit
vector of $q$ is defined as $\hat z^\prime$, and the quark plane is
formed by the $\hat z^\prime$ and the $\hat z$ axes of Collins-Soper
frame. The angular coefficients $A_i$ in Eq.~(\ref{eq:eq3}) can be
expressed in term of $\theta_1$ and $\phi_1$ as follows:
\begin{eqnarray}
A_0 =  \langle\sin^2\theta_1\rangle, ~~~
A_1 = \frac{1}{2} \langle\sin 2\theta_1\cos \phi_1\rangle, ~~~
A_2 =  \langle\sin^2\theta_1 \cos 2\phi_1\rangle,
\label{eq:eq4}
\end{eqnarray}
where the $\theta_1$ and $\phi_1$ are the polar and azimuthal angles
of the natural quark axis $\hat z^\prime$ of the quark plane in the
Collins-Soper frame.

Up to NLO ($\mathcal{O}(\alpha_S)$) in pQCD, the quark plane coincides
with the hadron plane and $\phi_1=0$. Therefore $A_0=A_2$ or
$1-\lambda-2\nu=0$, i.e., the L-T relation is satisfied. Higher order
pQCD processes allow the quark plane to deviate from the hadron plane,
i.e., $\phi_1 \neq 0$. This acoplanarity effect leads to a violation
of the L-T relation. For a nonzero $\phi_1$, Eq.~(\ref{eq:eq4}) shows
that $A_2 < A_0$. Therefore, when the L-T relation is violated, $A_0$
must be greater than $A_2$ or, equivalently, $1 - \lambda - 2\nu
>0$. This expectation of $1 - \lambda - 2\nu >0$ in this geometric
approach agrees with the results of NNLO pQCD calculations shown in
Fig.~\ref{fig2}. The geometric approach offers a simple and intuitive
interpretation for this result.

Furthermore the sign of $\mu$ could be either positive or negative,
depending on which parton and hadron the gluon is emitted
from~\cite{chang17,chang18}. Hence, one expects some cancellation
effects for $\mu$ among contributions from various processes. Each
process is weighted by the corresponding parton density
distributions. At mid-rapidity, the momentum fraction carried by the
beam parton is comparable to that of the target parton. Therefore, the
weighting factors for various processes are of similar magnitude and
the cancellation effect could be very significant, resulting in a
small value of $\mu$.

\section {Summary}

We have presented a comparison of the measurements of the angular
parameters $A_0$ and $A_0-A_2$ of the $Z$ production from the CMS
experiment as well as $\lambda$, $\mu$, $\nu$ and $1-\lambda-2\nu$ of
the D-Y process from the fixed-target E615 experiment with the
corresponding results from the NLO and NNLO pQCD
calculations. Qualitatively the transverse momentum dependence of
measured angular parameters could be described by pQCD. The L-T
violation part $A_0-A_2$ or $1-\lambda-2\nu$ remains zero in the NLO
pQCD calculation and turns positive in NNLO pQCD. The measurement of
$A_0$ and $A_2$ coefficients in $Z$ plus single-jet or multi-jet
events would provide valuable insights on the origin of the violation
of the L-T relation and could be used an an index in discriminating
the intrinsic property of high-$q_T$ jets.

Within the geometric picture, the occurrence of acoplanarity between
the quark plane and the hadron plane ($\phi_1 \neq 0$), for the pQCD
processes beyond NLO offers an interpretation of a violation of the
L-T relation. The predicted positive value of $1-\lambda-2\nu$, or
$A_0>A_2$ when $\phi_1$ is nonzero, is consistent with the NNLO pQCD
results.


\begin{thebibliography}{9}

\bibitem{peng14} J.~C.~Peng and J.~W.~Qiu, Prog. Part. Nucl. Phys.  {\bf
  76}, 43 (2014).

\bibitem{lam80} C.~S.~Lam and W.~K.~Tung, Phys. Rev. D {\bf 21}, 2712
  (1980).

\bibitem{cms} CMS Collaboration, V.~Khachatryan {\em et al.}, Phys.
Lett. B {\bf 750}, 154 (2015).

\bibitem{atlas} ATLAS Collaboration, G. Aad {\em et al.},
JHEP {\bf 08}, 159 (2016).

\bibitem{Gauld:2017tww} R.~Gauld, A.~Gehrmann-De Ridder, T.~Gehrmann,
  E.~W.~N.~Glover and A.~Huss,
  JHEP {\bf 1711}, 003 (2017).

\bibitem{Lambertsen:2016wgj} M.~Lambertsen and W.~Vogelsang,
  Phys.\ Rev.\ D {\bf 93}, 114013 (2016).

\bibitem{chang18}
  W.~C.~Chang, R.~E.~McClellan, J.~C.~Peng and O.~Teryaev,
  Phys.\ Rev.\ D {\bf 99}, 014032 (2019).

\bibitem{falciano86} NA10 Collaboration, S.~Falciano {\em et al.},
  Z. Phys. C {\bf 31}, 513 (1986); M.~Guanziroli {\em et al.},
  Z. Phys. C {\bf 37}, 545 (1988).

\bibitem{conway} E615 Collaboration, J.~S.~Conway {\em et al.}, Phys.
  Rev. D {\bf 39}, 92 (1989); J.~G.~Heinrich {\em et al.}, Phys. Rev.
  D {\bf 44}, 1909 (1991).

\bibitem{e866} E866/NuSea Collaboration, R.~S.~Towell {\em et al.},
  Phys.\ Rev.\ D {\bf 64}, 052002 (2001).

\bibitem{boer99} D. Boer, Phys. Rev. D {\bf 60}, 014012 (1999).

\bibitem{peng19} 
  J.~C.~Peng, W.~C.~Chang, R.~E.~McClellan and O.~Teryaev,
  Phys.\ Lett.\ B {\bf 797}, 134895 (2019).

\bibitem{peng16} J.~C.~Peng, W.~C.~Chang, R.~E.~McClellan, and
  O.~Teryaev, Phys. Lett. B {\bf 758}, 384 (2016).

\bibitem{chang17}
  W.~C.~Chang, R.~E.~McClellan, J.~C.~Peng and O.~Teryaev,
  Phys.\ Rev.\ D {\bf 96}, 054020 (2017).

\bibitem{peng18} 
  J.~C.~Peng, D.~Boer, W.~C.~Chang, R.~E.~McClellan and O.~Teryaev,
  Phys. Lett. B {\bf 789}, 356 (2019).

\bibitem{cs} J. C. Collins and D. E. Soper, Phys. Rev. {\bf D16},
    2219 (1977).

\bibitem{collins} J.~C.~Collins, Phys. Rev. Lett. {\bf 42}, 291 (1979).

\bibitem{thews} R.~L.~Thews, Phys. Rev. Lett. {\bf 43}, 987 (1979).

\bibitem{lindfors} J.~Lindfors, Phys. Scr. {\bf 20}, 19 (1979).

\bibitem{catani07} S. Catani and M. Grazzini, Phys. Rev. Lett.
{\bf 98}, 222002 (2007).

\bibitem{catani09} S. Catani {\em et al.}, Phys. Rev. Lett.
{\bf 103}, 082001 (2009).


\end{thebibliography}
\end{document}